\RequirePackage{fix-cm}
\documentclass[twocolumn]{svjour3}                     
\smartqed  
\usepackage{natbib}
\usepackage{graphicx}
\usepackage{algorithm}
\usepackage{algorithmic} 
\usepackage{amssymb}
\usepackage{graphicx}
\usepackage{paralist}
\usepackage{url}
\usepackage{subfig}
\usepackage{array}
\usepackage{color} 
\newcolumntype{L}[1]{>{\raggedright\let\newline\\\arraybackslash\hspace{0pt}}m{#1}}
\newcolumntype{C}[1]{>{\centering\let\newline\\\arraybackslash\hspace{0pt}}m{#1}}
\newcolumntype{R}[1]{>{\raggedleft\let\newline\\\arraybackslash\hspace{0pt}}m{#1}}
\usepackage{tabularx}
\usepackage{booktabs}
\journalname{Social Network Analysis and Mining}

\graphicspath{ {./JournalFigures/} }

\begin{document}

\title{Identifying Community Structures in Dynamic Networks\thanks{Research at University of Central Florida was supported by NSF award IIS-0845159.  Sandia National Laboratories is a multi-program laboratory managed and operated by Sandia Corporation, a wholly owned subsidiary of Lockheed Martin Corporation, for the U.S. Department of Energy's National Nuclear Security Administration under contract DE-AC04-94AL85000.}}
		

\author{\centering Hamidreza Alvari$^{1}$ \and Alireza Hajibagheri$^{1}$ \and Gita Sukthankar$^{1}$ \and Kiran Lakkaraju$^{2}$}

\authorrunning{Alvari et al\.} 

\institute{
$^{1}$ University of Central Florida, Orlando, Florida,\\
\url{{halvari,hajibagheri,gitars}@eecs.ucf.edu}\\
$^{2}$ Sandia National Labs, Albuquerque, New Mexico,\\
\url{klakkar@sandia.gov}
}

\date{Received: December 11, 2015. Accepted: August 31, 2016.}

\maketitle

\begin{abstract}
Most real-world social networks are inherently dynamic, composed of communities that are constantly changing in membership.   To track these evolving communities, we need dynamic community detection techniques.   This article evaluates the performance of a set of game theoretic approaches for identifying communities in dynamic networks.   Our method, D-GT (\textbf{D}ynamic \textbf{G}ame \textbf{T}heoretic community detection), models each network node as a rational agent who periodically plays a community membership game with its neighbors.  During game play, nodes seek to maximize their local utility by joining or leaving the communities of network neighbors.   The community structure emerges after the game reaches a Nash equilibrium.  Compared to the benchmark community detection methods, D-GT more accurately predicts the number of communities and finds community assignments with a higher normalized mutual information, while retaining a good modularity.
\end{abstract}

\section{Introduction}
\label{sec:intro}
The natural flux of people's changing social ties and interests generates a dynamic social network.  This invisible network can be observed by capturing daily or weekly snapshots of user activities on social media platforms and massively multiplayer online games (MMOGs).  It is informative to study changes in the network at the \textit{community level}, as well as the individual level.  Communities are emergent groups that are created as people form highly connected subnetworks with their families, co-workers, and friends.  Often communities are formed by participants with the same goals, interests, or a geographic location.  For instance, in MMOGs, network communities may emerge from guilds of players with common economic interests or alliances who share strategic goals.  As the network changes, user groups can grow, shrink, or disappear, causing drastic changes in the total number of network communities.

Community detection can help us understand the hidden social structure of the user populations, but the dynamic aspect of networks can pose problems for standard algorithms.  Our method uses stochastic optimization to find the best community structure, assuming that the nodes are modeled as rational players who seek to maximize their personal utility while playing a \textit{community membership game} with neighboring nodes.  In this game, the active agent decides to join or leave different communities; agents receive benefits from being part of the same community as their network neighbors but are penalized for joining too many communities.  The Nash equilibrium of the current game corresponds to the community structure of the current snapshot.  As the network evolves, agents usually find it advantageous to modify their community membership strategy.  This article examines the performance of varying the amount of information propagated from prior snapshots.  Even in dynamic networks, there are many nodes that retain the same community membership or rejoin their former communities.  Thus propagating information from previous snapshots can provide more favorable initialization conditions for the stochastic optimization procedure.

Much of the power of the D-GT framework (originally introduced in \citep{6921567}) lies in its potential for customization; however, without guidance end users can be overwhelmed by myriad choices.  Our aim in this article is to present a comprehensive analysis of the algorithm's performance in common scenarios; while retaining the same basic game-theoretic model, we evaluate different variations of our procedure.   First, we examine the performance of different utility functions, a similarity-based utility function~\citep{alvari2011detecting} vs.\ the use of a personalized modularity function~\citep{1842566}.  Then we compare different initialization approaches in which the following information is propagated: 1) no information (\textbf{D-GTS}); 2) the union of the community membership information over all snapshots (\textbf{D-GT}); 3) community membership information from the previous snapshot (\textbf{D-GTP}); 4) ground truth information for a small seed set (\textbf{D-GTG}).   Our game-theoretic model is robust to minor changes in the procedure and most variants outperform the benchmarks.  

Our experiments were conducted on networks created from different dynamic processes: internet routers (AS-Oregon Graph, the AS-Internet Routers Graph), shifting organization structure (Enron Email dataset), citation graphs from arXiv (hep-ph), and player interactions in massively multiplayer online games (Travian Messages, Travian Trades).  Results were compared against five other methods: LabelRankT, iLCD, OSLOM, InfoMap and Louvain in terms of normalized mutual information (NMI), modularity, and the number of detected communities.  The next section presents an over-view of related work on community detection in dynamic networks.  Section~\ref{sec:method} describes our problem formulation and our proposed method (\textbf{D}ynamic \textbf{G}ame \textbf{T}heoretic community detection).  Experimental results are provided in Section~\ref{sec:results}, before we conclude the article (Section~\ref{sec:conclusion}).  

\section{Related Work}
\label{sec:related}
The problem of community detection has been widely studied in the literature compared to the other research areas in the social networks~\citep{yang2010modeling, beigi2016exploiting,leskovec2010predicting, beigi2016signed,beigi2016overview}. In particular, in static networks, community detection has appeared in multiple disciplines including sociology and computer science.  This has yielded a diverse set of approaches ranging from traditional network structure based algorithms~\citep{Girvan2002,Newman2006,newman2004}, optimization techniques~\citep{alvari2011detecting,1842566}, label propagation~\citep{raghavan2007near,xie2011community,2012towards}, propinquity~\citep{Zhang:2009:PCD:1557019.1557127} and information diffusion~\citep{hajibagheri2013modeling,hajibagheri2012community}.  Detecting community structure in dynamic networks, on the other hand, has attracted less research attention due to the complexity of the problem and dearth of good datasets. There are some community detection algorithms originally design-ed for static networks that continue to perform well in dynamic datasets. For instance, Lancichinetti et al.'s OSLOM (Order Statistics Local Optimization Method) works on single snapshots but also benefits from information from previous network partitions.  Like D-GT, OSLOM's optimization procedure can be initialized with the partition from the previous snapshot; it aims to optimize cluster significance with respect to a global null model~\citep{lancichinetti2011finding}.  We use this method as one of our benchmarks, along with two other static community detection algorithms, Louvain \citep{blondel2008fast} and InfoMap~\citep{rosvall2008maps}.   These algorithms perform well on many static community detection problems and have the benefit of being fast to compute on a single network snapshot.

Other network properties have also been used to perform dynamic community detection; for instance Hui et al.~\citeyearpar{hui2007distributed} proposed a distributed method for community detection in which modularity was used as a measure instead of the objective function.  QCA (Quick Community Adaptation) is a modularity-based approach that focuses explicitly on the changes in the network structure, rather than recomputing community structure from scratch at each time step~\citep{10.1371/journal.pone.0091431}.  In this article, we evaluate the use of personal modularity as an alternative gain function to neighborhood similarity.

Some studies have focused on studying the evolution of communities over time.  For instance, \cite{hopcroft2004tracking} identified subsets of nodes, ``natural communities", that were stable to small perturbations of the input data.   Communities detected in later snapshots were matched to earlier snapshots using the natural community tree structure. Palla et al.~\citeyearpar{palla2009social} proposed an innovative method for detecting communities in dynamic networks based on the k-clique percolation technique; in their approach, communities are defined as adjacent k-cliques, that share $k-1$ nodes. 
	
Machine learning has also been employed to model changes in community structure; for instance, \cite{DBLP:conf/asunam/TakaffoliRZ14} predict transitions in community structure by learning supervised machine learning classifiers.  This requires data on past transitions to train the classifiers, which limits its applicability to certain datasets.  \cite{sun2007graphscope} adopt a data mining approach to detect clusters on time-evolving graphs; community discovery and change detection are performed using the minimum description length (MDL) paradigm.

Rather than independently detecting communities at each snapshot and matching them, another option is to make a local decision to add nodes to existing communities when new edges appear in the network.  One of our benchmarks, iLCD (intrinsic Longitudinal Community Detection)~\citep{cazabet2010detection}, updates the community structure of the network based on time-stamped sets of edges.  Nodes are added to communities if its mean number of second neighbors and robust second neighbors exceeds the current average for the community.  However, this model is limited to certain types of network changes, and cannot handle interconnected pairs of nodes being simultaneously added or the removal of edges.

Optimization can be used to identify minimum cost community assignments in dynamic graphs.  FacetNet \citep{lin2008facetnet} is a framework for analyzing communities in dynamic networks based on an optimization of snapshot costs. It is guaranteed to converge to a local optimal solution; however, its convergence speed is slow, and it needs to be initialized with the number of communities which is usually unknown in practice.  \cite{6573961} modeled dynamic community detection as a multi-objective optimization problem.  Their approach is parameter free and uses evolutionary clustering to optimize a dual objective function.  The first objective selects for highly modular structures at the current time step, and the second minimizes the differences between community structures in the current and previous time steps. D-GT also uses a stochastic optimization procedure, but all of the agents individually optimize their utilities based on local network information.

The Markov Cluster Algorithm (MCL)~\citep{van2000cluster} identifies graph clusters by computing the probabilities of random walks through the graph; flow simulations are performed by alternating expansion (matrix squaring) with inflation operations (Hadamard powers).  Due to its mathematical simplicity, it is popular for community detection in many domains but is slow and often overestimates the number of communities in the dataset.  Regularized-MCL~\citep{satuluri2009scalable} is a variant of MCL that prevents overfitting by taking into account neighbor flows.  It can also be used within a multi-level framework (Multi-level Regularized MCL) to speed up computation by executing a sequence of coarsening operations on the graph before executing R-MCL.

LabelRankT~\citep{xie2013labelrankt}, shares some similarities with the MCL techniques described above while improving upon them in several ways; it is a label propagation approach in which the inflation operation is applied to the label distribution matrix rather than to the adjacency matrix.  Each node requires only local information during propagation making it more scalable than MCL and amenable to parallelization. Due to LabelRankT's strong performance and good implementation, it was selected as the best label propagation benchmark for our work.

Our proposed method (D-GT) attempts to simulate the decision-making process of the individuals creating the communities, rather than focusing on statistical correlations between labels of neighboring nodes. We believe that exploiting game theory for dynamic community detection yields more realistic, fine-grained communities since intrinsically game theory is a good representation for expressing the behavior of individuals and strategic interactions among them~\citep{adjeroh2007game}. 

In previous work, we have demonstrated the success of game-theoretic approaches in static community detection across several domains, including detecting guilds in massively multiplayer online games~\citep{sbp14} and predicting trust between users on e-commerce sites \citep{gbeigi}.  Many of these domains featured overlapping communities in static networks~\citep{alvari2011detecting,alvari2013}; however in this article the datasets are dynamic, but not overlapping.  D-GT makes a hard assignment at the termination of the stochastic optimization procedure by selecting the community assignment with the highest utility function.  In this article, we investigate the performance of different gain functions and initialization procedures, on a variety of evaluation metrics (modularity, NMI, number of detected communities). The strength of the D-GT framework is its versatility; our results show that substantial performance improvements can be achieved by customizing it for the problem at hand.

\section{Method}
\label{sec:method}
First, we provide the formal definition of the problem and the notations used throughout the paper. Given snapshots $\textbf{T} = \{T^t \mid \forall t, t=1,...,M \}$ of a dynamic network and their corresponding underlying graphs $\textbf{G$^t$} = (V^t,E^t)$, with $n^t = |V^t|$ vertices and $m^t = |E^t|$ edges, where t=1,...,M, we aim to detect community structure $\textbf{C}=\{C^t \mid \forall t, t=1,...,M \}$ of the network.  Table~\ref{tb:symb} shows the symbols and definitions used throughout the paper. 

We leverage a dynamic agent-based model to detect communities by optimizing each user's utility through a stochastic search process~\citep{alvari2011detecting}.  In this article, we evaluate four different variants of the procedure: D-GT (\textbf{D}ynamic \textbf{G}ame \textbf{T}heoretic community detection), D-GTP (\textbf{D-GT} with passing one \textbf{P}revious Snapshot) D-GTS (\textbf{D-GT} with \textbf{S}eparate Runs) and D-GTG (\textbf{D-GT} with passing \textbf{G}round Truth).

\begin{table}
 \caption{Definition of Symbols.}
\centering
 \label{tb:symb}
 \begin{tabular}{l|l}
   \toprule
   \textbf{Symbol} & \textbf{Definition}\\\hline
  \textbf{T} & set of snapshots\\
   \textbf{C} & set of communities\\
   \textbf{G$^t$} & graph of $t$-th snapshot with no self-edges\\
  $C^t$ & community structure of $t$-th snapshot\\	
   $C_k^t$ & $k$-th community in $C^t$\\
   $m^t$, $n^t$ & number of edges and vertices of $G^t$ \\
   \textbf{A$^t$} & adjacency matrix of $G^t$\\  
  \textbf{S$^t$} & profile of strategies of $t$-th snapshot\\
  $s_{i}^t$ & $i$-th agent's strategy in $t$-th snapshot\\
  $g_{i}^t$ & $i$-th agent's gain function in $t$-th snapshot\\
  $l_{i}^t$ & $i$-th agent's loss function in $t$-th snapshot\\
  $u_{i}^t$ & $i$-th agent's utility function in $t$-th snapshot\\
  $c_{ij}^t$ & similarity between $i$-th and $j$-th agents\\
    & in the $t$-th snapshot\\\bottomrule
 \end{tabular}
\end{table}

\begin{table}
 \caption{Definition of Possible Actions.}
\centering
 \label{tb:act}
 \begin{tabular}{l|l}
   \toprule
   \textbf{Action} & \textbf{Definition}\\\hline
  \textbf{Join} &  add a new label to $s_i^t$\\
  \textbf{Leave} & remove a label from $s_i^t$\\
  \textbf{Switch} & remove a label  from $s_i^t$ and add a new one\\
  \textbf{No operation} & no action is performed\\\bottomrule
 \end{tabular}
\end{table}

\subsection{Dynamic Game Theory (D-GT)}
In this section, we present the framework for D-GT.  We treat the process of community detection as an iterative game performed in a dynamic multi-agent environment in which each node of the underlying graph is a selfish agent who decides to maximize its total utility $u_i$. Note that hereafter we use the terms \textit{node}, \textit{user}, and \textit{agent} interchangeably. The terms \textit{pool}, \textit{list}, or \textit{set} of agents are used to denote the set of nodes maintained during each game.

During the community formation game, each agent assesses whether taking an action ({\it join}, {\it switch}, or {\it leave}) (Table~\ref{tb:act}) will increase its utility.  An agent \textbf{joins} a new community $c^t \subseteq C^t$ by adding its label to $s_i^t$.  It then gains utility $u_{Join}^t$ and its community membership changes:
\begin{equation}\label{1}
s_i^t \leftarrow s_i^t \cup \{c^t\}.
\end{equation}

It may \textbf{leave} one of its own communities, say $c'^t$ by removing its label from $s_i^t$ which results in utility $u_{Leave}^t$ and changes the community membership as follows:
\begin{equation}\label{2}
s_i^t \leftarrow s_i^t / \{c'^t\}.
\end{equation}
The agent can also simultaneously switch communities:
\begin{equation}\label{3}
s_i^t \leftarrow s_i^t / \{c'^t\},
s_i^t \leftarrow s_i^t \cup \{c^t\}.
\end{equation}
Even though the \textbf{switch} action is not strictly necessary, having this additional action speeds up the convergence of the stochastic search process.

The new utility $u'^t_i$ for this agent is updated as follows:
\begin{equation}\label{4}
u'^t_i \leftarrow \max\{u_{Join}^t,u_{Leave}^t,u_{Switch}^t,u_{noOp}^t\}.
\end{equation}

\noindent To reduce computation time, only communities containing the agent's nearest network neighbors are considered as candidates for the join/switch operation, and we independently identify the best communities for the join and leave operations.  If no action improves the agent's utility, it does not change its strategy ({\it no operation}).

The set of all communities at the $t$-th snapshot is denoted by $C^t$. We define a strategy profile $S^t = (s_1^t, s_2^t, ..., s_n^t)$ which represents the set of all strategies of all agents, where $s_i^t \subseteq C^t$ denotes the strategy of agent $i$, i.e. the set of its labels at snapshot $t$.  In our framework, for each snapshot, the best response strategy of an agent $i$ with respect to strategies $S_{-i}^t$ of other agents is calculated as:

\begin{equation}\label{5}
\arg \max_{s_i^t \subseteq C^t} u^t_i(S_{-i}^t,s_i^t)
\end{equation}
Agents are selected randomly, without replacement, until all the agents have had the opportunity to play the community formation game in order to guarantee adequate exploration of the strategy search space. The utility function for each agent is calculated by combining the benefit of its community memberships, based on a gain function and subtracting losses incurred:

\begin{equation}\label{eq:utility}
	u^t_i(S_{-i}^t,s_i^t)=g^t_i(S_{-i}^t,s_i^t) - l^t_i(S_{-i}^t,s_i^t),
\end{equation}

We have experimented with two variants on the gain function for agent $i$.\footnote{We employ the notation for directed graphs, although it is straightforward to generalize to undirected graphs by ignoring the \textit{in/out} superscripts.}
The first gain function is based on \textbf{similarity} between agents:
{
\small
\begin{equation}\label{eq:gain1}
	g^t_i(S^t)=\frac{1}{m^t}\sum_{k \in s_i^t} \sum_{j \in C_k^t,j\neq i}c_{ij}^t.
\end{equation}
}

Here, we use neighborhood similarity to quantify the structural equivalence between users at time $t$:

\begin{equation}\label{eq:sim}
c_{ij}^t=\left\{
\begin{array}{c l}      
	w_{ij}^t(1-d_i^{in,t}d_j^{out,t}/2m^t) & A_{ij}^t = 1 , w_{ij}^t >= 1\\
   w_{ij}^t/n^t & A_{ij}^t = 0 , w_{ij}^t >= 1\\
d_i^{in,t}d_j^{out,t}/4m^t & A_{ij}^t = 1 , w_{ij}^t = 0\\
-d_i^{in,t}d_j^{out,t}/4m^t & A_{ij}^t = 0 , w_{ij}^t = 0
\end{array}\right.
\end{equation}

\noindent $w_{ij}^t$ is defined as the number of common neighbors possessed by nodes $i$ and $j$, where common neighbors are nodes with direct in-edges from both $i$ and $j$. $d_i^{in,t}$ and $d_i^{out,t}$ are the in- and out-degrees of node $i$ at snapshot $t$.  $m^t$ and $n^t$ are the number of edges and nodes respectively and primarily serve as normalization constants.  Note that $c_{ij}^t$ assumes its highest value when two nodes have at least one common neighbor and are also directly connected, i.e. $A_{ij}^t=1$. Hence agents playing the community formation game benefit from joining communities containing connected nodes with many common neighbors.

The second gain function measures the personalized \textbf{modularity} of the $i$-th agent:

{\scriptsize
\begin{equation}\label{eq:gain2}
	g_i^t(S^t)=\frac{1}{2m^t}\sum_{k \in s_i^t} \sum_{j \in C_k^t,j\neq i} \sum_{k' \in s_j^t} (A_{ij}^t\delta(i,j)-\frac{d_i^{in,t}d_j^{out,t}}{2m^t}|k \cap k'|),
\end{equation}
}

\noindent where $k \in s_i^t$ and $k' \in s_j^t$ refer to the community labels at snapshot $t$ that agent $i$ and $j$ belong to respectively; $\delta(i,j)$ is an indicator function that is 1 when $i$ and $j$ are members of the same community. More intuitively, this gain function explains how well the $i$-th agent fits the communities it belongs to, compared to a randomly assigned community.

Similar to what happens in real life, we also consider the loss function $l_i^t$ for each agent, which is linear in the number of labels each agent has, This can be used to model the intrinsic communication overhead of belonging to multiple communities and prevents the agents from indiscriminately joining every available community. Therefore we define the following loss function for agent $i$:

\begin{equation}\label{eq:loss}
	l_i^t(S_{-i}^t,s_i^t)=\frac{|s_i^t|}{m^t}.
\end{equation}

Here $s_i^t = \{1,2,...,k\}$ is the set of labels at snapshot $t$ which agent $i$ belongs to.

In our framework, the best response strategy of the agent $i$ with respect to strategies $S_{-i}^t$ of other agents is calculated by:
\begin{equation}\label{6}
\arg \max_{s'^t_i \subseteq C^t} g^t_i(S_{-i}^t,s'^t_i) - l^t_i(S_{-i}^t,s'^t_i).
\end{equation}

The strategy profile $S^t$ forms a pure Nash equilibrium of the community formation game if no agent can unilaterally improve its own utility by changing its strategy:
\begin{equation}\label{7}
 \forall i, s'^t_i \neq s_i^t, u^t_i(S_{-i}^t,s'^t_i) \leq u^t_i(S_{-i}^t,s_i^t).
\end{equation} 

A local equilibrium is reached if all agents play their local optimal strategies:
\begin{equation}\label{8}
\forall i, s'^t_i \in ls(s_i^t), u_i^t(S_{-i}^t,s'^t_i) \leq u_i^t(S_{-i}^t,s_i^t).
\end{equation} 
Here $ls(s_i^t)$ refers to local strategy space of agent $i$, which is the set of \emph{possible} label sets it can obtain by performing the actions defined earlier.

\subsection{Existence of Equilibria}
The evolving community structure present in dynamic networks is accounted for by propagating strategies from prior snapshots.  D-GT uses the same community formation game described in \cite{alvari2013}.  Since strategy propagation only changes the initialization conditions, the same proof of the existence of Nash equilibria in the community formation game applies to D-GT as well; we summarize the argument below.

To see when a certain game has Nash equilibria, recall that potential games are a general class of games that permit pure Nash equilibria~\citep{AlgGameTheory}. For any finite game, there exists a potential function $\Theta$ defined on the strategy profile $S$ of the agents that maps this profile to some real values. This function must validate the following condition: 

\begin{eqnarray}\label{proof1}
\forall i, \Theta(S^t)- \Theta(S_{-i}^t,s'^t_i) =  u_i^t(S_{-i}^t,s'^t_i) - u_i^t(S^t). \nonumber \\
\end{eqnarray} 

Equivalently, if the current strategy profile of the game is $S^t$ and the agent $i$ switches from strategy $s_i^t$ to $s'^t_i$, the potential function exactly mirrors the changes in the agent utility. It is not hard to see that a game has at most one potential function. A game that does possess a potential function is called a potential game. Consequently we have the following theorem:

\begin{description}
\item[\textbf{Theorem 1.}] Every potential game has at least one pure Nash equilibrium, namely the strategy profile $S$ that minimizes $\Theta(S)$ ~\citep{AlgGameTheory}. 

\item[\textbf{Proof.}] Let $\Theta$ be a potential function for this game and let $S$ be a pure strategy profile minimizing $\Theta(S)$. 
Consider any action performed by player $i$ that results in a new strategy profile $S'$. By assumption,  $\Theta(S' ) \geq \Theta(S)$ and by the definition of a potential function, $u_i (S' )-u_i (S)=\Theta(S)-\Theta(S' )$. Thus the utility of agent $i$ cannot increase from this move and hence $S$ is stable~\citep{1842566}.
\end{description}

Now we provide a sufficient condition to prove our community formation game as a potential game and thus address the existence of the Nash equilibrium. First we have the following definition~\citep{1842566}:

\begin{description}
\item[\textbf{Definition}] [\textmd{\textbf{Locally linear function}}]\textbf{.} A set of functions $\{f_i, 1 \leq i \leq n\}$ is locally linear with locality factor $\rho$ if for every strategy profile S the following condition holds: 
\end{description}
\begin{eqnarray}\label{proof2}
\forall i, f_i(S_{-i},s'_i)-f_i(S) = \rho(f(S_{-i})-f(S)). \nonumber \\
\end{eqnarray} 
where $f(.)= \sum_{i \in [n]} f_i(.)$. According to Theorem 2, if we show that our gain and loss functions are locally linear, then we can prove the existence of Nash equilibrium in our framework.

\begin{description}
\item[\textbf{Theorem 2.}] Let  $\{g_i,   1 \leq i \leq n\}$ and $\{l_i,   1 \leq i \leq n\}$ be the sets of gain and loss functions of a community formation game. If these sets are locally linear functions with linear factors $\rho_G$ and $\rho_L$, then the community formation game is a potential game~\citep{AlgGameTheory}.

\item[\textbf{Proof.}] We define a potential function as $\Theta(S^t)= \rho_ll(S^t)-\rho_gg(S^t)$ and assume that agent $i$ who changes its strategy from $s_i^t$ to $s'^t_i$. Based on the definitions of locally linear functions and the utility functions $u_i^t (.)$, we have $ \Theta(S^t)-\Theta(S_{-i}^t,s'^t_i )= u_i^t (S_{-i}^t,s'^t_i ) - u_i^t (S^t)$. Therefore, the community formation game is a potential game~\citep{1842566}. 
\end{description}

\subsection{Algorithm}
An overview of the D-GT framework is shown in Algorithm~\ref{alg:alg1}.  For every snapshot of the network, a set of agents, one representing each node in the graph, is created to play the community formation game.  The community structure is initialized either with a set of singleton communities or with communities passed from previous snapshots.  During game play, an agent is randomly selected (without replacement) from the pool; it selects an action (join, leave, switch, or no op) by calculating the strategy that yields the highest utility.  After the agent plays, the community structure is updated. 

\begin{figure}
 \centering
 {\includegraphics[width=0.9\columnwidth]{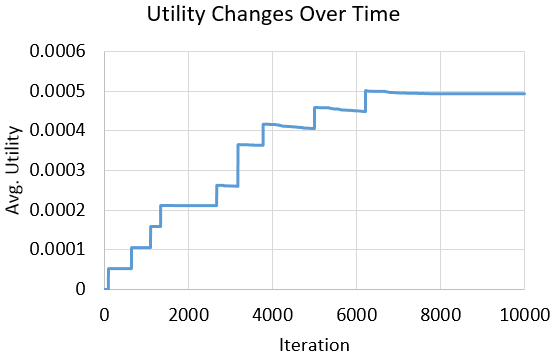}}
 \caption{Change in average utility summed over all nodes vs.\ iteration for the Travian-Trades dataset (one snapshot with 964 nodes). The algorithm converges after 6680 iterations which requires 2.8 seconds to complete.}
  \label{fig:utilitytime}
\end{figure}

The game is played until the number of agents changing their play strategy between permutations falls below the threshold, or the maximum iteration is reached.   Empirically, we have discovered that $8n$ is a good iteration limit with a threshold of 5\%.  Thus if there are 1000 nodes in a network snapshot, the community formation game is played until fewer than 50 nodes change strategies or to the maximum of 8000 games.  Figure~\ref{fig:utilitytime} shows an example of the convergence in utility vs.\ iteration.  

The outer loop of the algorithm requires iterating over $M$ graph snapshots, and the inner loop requires performing a maximum of $8n$ iterations of the community formation game that, in the worst case, requires considering $n$ community choices.  Thus, the overall time complexity of D-GT is $O(Mn^2)$.

D-GT maintains a candidate set of multiple community assignments per agent until the last iteration and then selects the assignment with the highest utility function as the final disjoint partition.  In this article, we evaluate several different variants of the procedure:

\begin{compactitem}
\item \textbf{D-GTS} (D-GT with Separate runs): this version does not employ any information from previous runs and hence is equivalent to a static community detection procedure.
\item \textbf{D-GTP} (D-GT passing one Previous Snapshot): rather than passing strategy profiles from all previous snapshots, we only initialize the community formation game with the structure from a single previous time slice. For each snapshot $T^k$, we initialized communities and agents to the existing information from $T^{k-1}$.
\item \textbf{D-GTG} (D-GT with passing Ground Truth): D-GTG leverages some ground truth information.  A select seed group of ground truth communities with predefined size is used to initialize communities for snapshot $T^k$; however we do not pass any discovered community structure to the following snapshots (similar to D-GTS). This variation cannot be used unless some of the agents' community membership is known in advance.
\end{compactitem}

\begin{algorithm}
    \caption{\textbf{D-GT} Community Formation Game}
    \label{alg:alg1}
    
    \begin{algorithmic}[1] 
\STATE \textbf{Input}: Snapshots $\textbf{T} = \{T^1,T^2,\ldots,T^M\}$
\STATE \textbf{Output:} Communities $\textbf{C} = \{C^1,C^2,\ldots,C^M\}$
\FORALL{$T^t \in \textbf{T}$}
    \STATE{Initialize $p \leftarrow 0$}
    \REPEAT
	\STATE{Initialize $q \leftarrow 0$}
	\FORALL{agents in randomized order}
	    \STATE{Select best action $a$ using Eqn.~\ref{5}}
	    \IF{$a = \mbox{``No operation''}$}
	    	\STATE{$q \leftarrow q + 1$}
	    \ELSE
	    	\STATE{Update $C^t$ according to $a$}
	    \ENDIF
	\ENDFOR
	\STATE{$p \leftarrow p + 1$}
    \UNTIL{$p>8$ or $q > \mbox{threshold } \theta$}
\ENDFOR
  \end{algorithmic}
  
\end{algorithm}

\section{Experimental Results}
\label{sec:results}

Algorithms were evaluated together on a system with 12G of RAM and Intel CPU 2.53 GHz, and all reported results were averaged over ten repetitions.  We compare D-GT with the following community detection baselines:

\begin{figure*}
 \centering
{\includegraphics[width=0.325\textwidth]{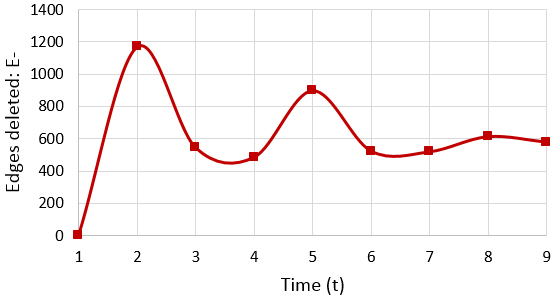}} 
   {\includegraphics[width=0.325\textwidth]{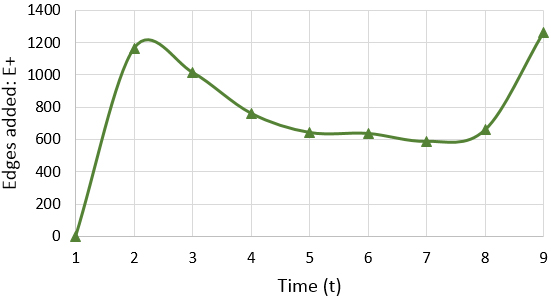}}
 {\includegraphics[width=0.325\textwidth]{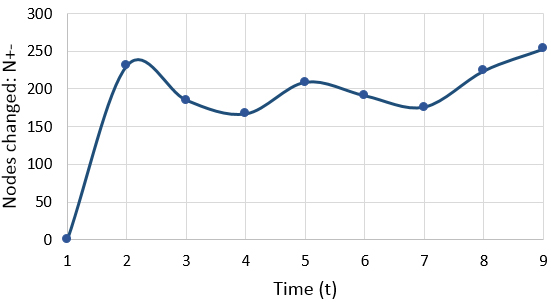}} \\
 \caption{The structural changes in the AS-Oregon dataset over 9 snapshots including the number of edges deleted $(E_-)$ and added $(E_+)$, as well as the number of nodes involved in changes $(N_{+ -})$. The community detection problem becomes more challenging when there are significant structural changes between snapshots.} 
  \label{fig:changes2}
\end{figure*}

\begin{figure*}
 \centering
  {\includegraphics[width=0.325\textwidth]{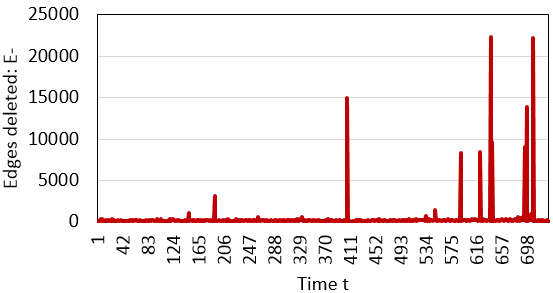}} 
   {\includegraphics[width=0.325\textwidth]{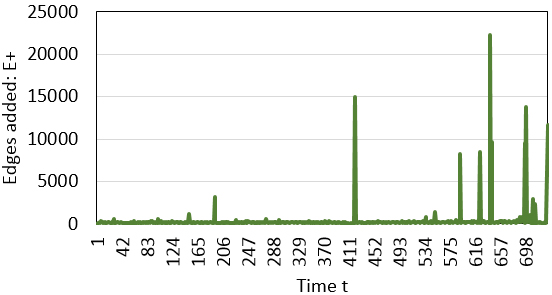}}
 {\includegraphics[width=0.325\textwidth]{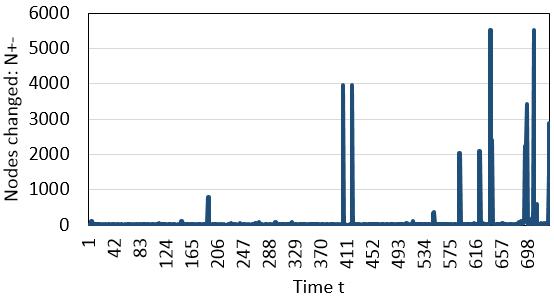}} \\

 \caption{The structural changes in the AS-Internet dataset over 733 snapshots.} 
  \label{fig:changes7}
\end{figure*}

\begin{figure*}
 \centering
{\includegraphics[width=0.325\textwidth]{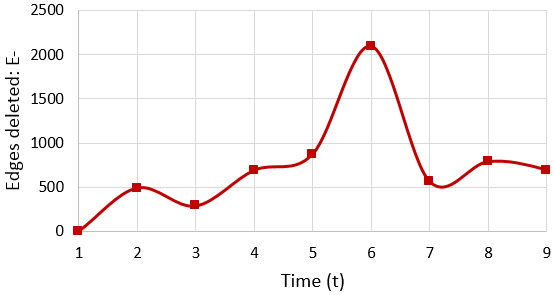}}
  {\includegraphics[width=0.325\textwidth]{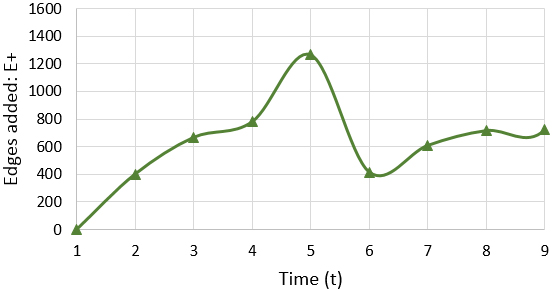}} 
 {\includegraphics[width=0.325\textwidth]{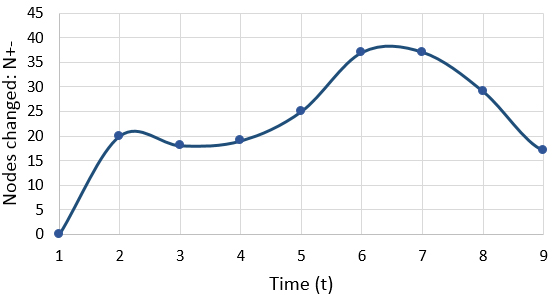}} \\
 \caption{The structural changes in the Enron dataset over 12 snapshots.} 
  \label{fig:changes3}
\end{figure*}

\begin{figure*}
 \centering
{\includegraphics[width=0.325\textwidth]{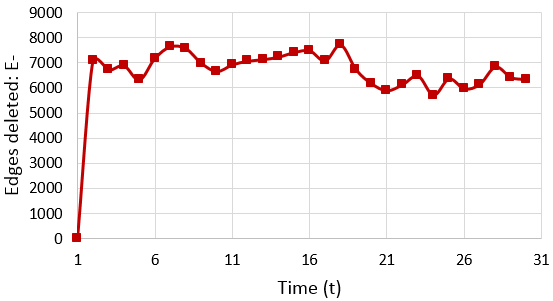}}
  {\includegraphics[width=0.325\textwidth]{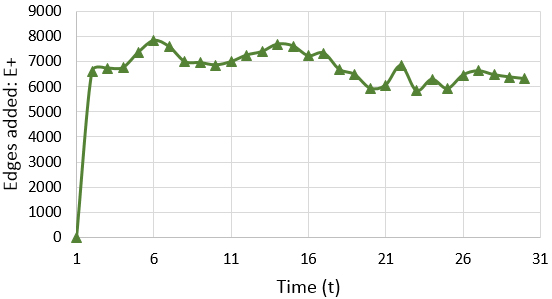}} 
 {\includegraphics[width=0.325\textwidth]{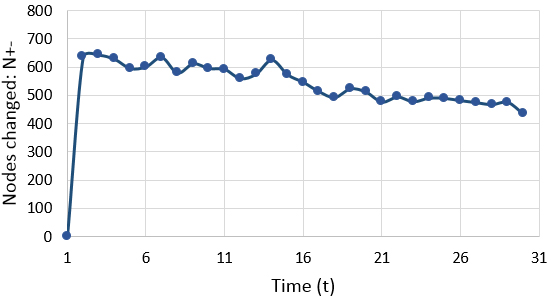}} \\
 \caption{The structural changes in the Travian Trades dataset over 30 snapshots.} 
  \label{fig:changes4}
\end{figure*}

\begin{figure*}
 \centering
   {\includegraphics[width=0.325\textwidth]{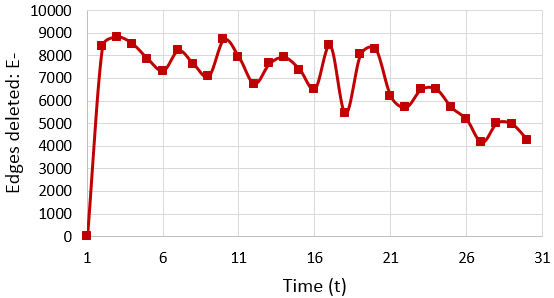}}
  {\includegraphics[width=0.325\textwidth]{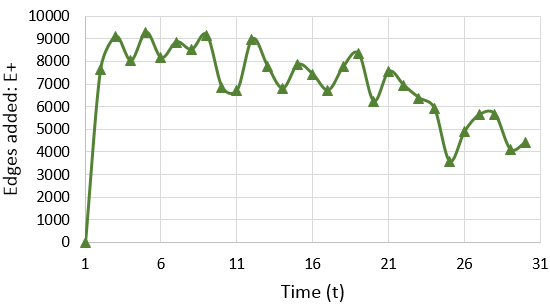}} 
 {\includegraphics[width=0.325\textwidth]{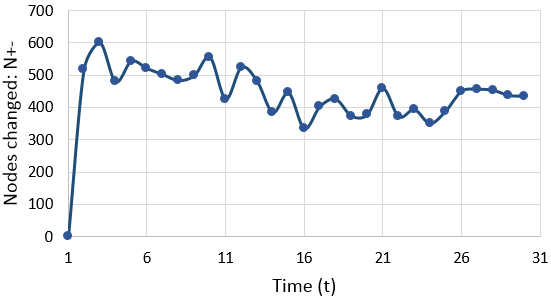}} \\
 \caption{The structural changes in the Travian Messages dataset over 30 snapshots.} 
  \label{fig:changes5}
\end{figure*}

\begin{figure*}
 \centering
  {\includegraphics[width=0.325\textwidth]{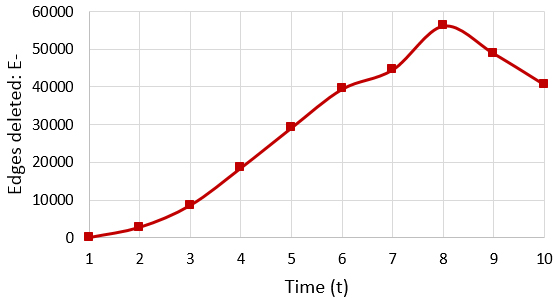}} 
   {\includegraphics[width=0.325\textwidth]{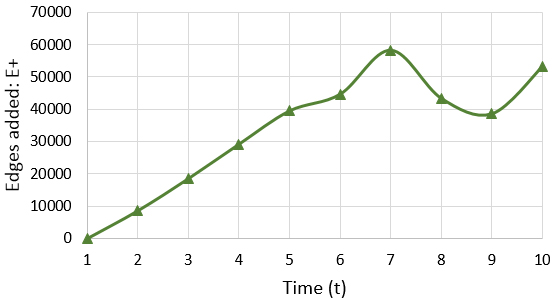}}
 {\includegraphics[width=0.325\textwidth]{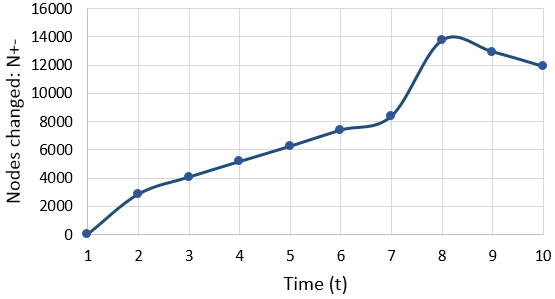}} \\
 \caption{The structural changes in hep-ph dataset over 10 snapshots.} 
  \label{fig:changes6}
\end{figure*}

\begin{itemize}
	\item \textbf{LabelRankT\footnote{https://sites.google.com/site/communitydetectionslpa}~\citep{xie2013labelrankt}.}  LabelRankT functions according to the generalized LabelRank, in which each node requires only local information during label propagation processing. Several parameters must be set before running the algorithm on the data; we used the best performing values reported in the original paper.
	
	\item \textbf{iLCD\footnote{http://www.cazabetremy.fr/iLCD.html}~\citep{cazabet2010detection}.} iLCD is another well known community detection approach for dynamic social networks which works by first adding edges and then merging the similar ones. It takes the dynamics of the network into account.  

	\item \textbf{OSLOM\footnote{http://www.oslom.org/software.htm}~\citep{lancichinetti2011finding}.}
The Order Statistics Local Optimization Method (OSLOM) is a versatile community detection algorithm that can handle most types of graph properties including edge directions and weights, overlapping communities, hierarchies and community dynamics.  It is based on the local optimization of a fitness function expressing the statistical significance of clusters with respect to random fluctuations.

\item\textbf{InfoMap\footnote{http://www.mapequation.org/code.html}~\citep{rosvall2008maps}.}
InfoMap is a static community detection method that calculates the probability flow of random walks and decomposes the network into modules by compressing a description of the flows.  Since this is a static algorithm, we run it separately on each snapshot.

\item\textbf{Louvain\footnote{https://sites.google.com/site/findcommunities}~\citep{blondel2008fast}.}
The Louvain method is a static community detection approach designed to optimize modularity using heuristics.
Small communities are found by optimizing modularity locally for all nodes.  Then each community is grouped into a single node, and the first step is repeated.  We run this algorithm separately on every network snapshot.
\end{itemize}

\subsection{Datasets}
To illustrate the strength and effectiveness of our approach, we selected some communication networks from the SNAP\footnote{http://snap.stanford.edu/} graph library as well as two networks (messages and trades) from a well-known multiplayer online game.  Statistics for the datasets are provided in Table~\ref{tab:dataset}, and description of the datasets is as follows:

\begin{table*}
\caption{Dataset Summary}
\label{tab:dataset}
\begin{center}
\begin{tabular}{lcccccc}
\toprule
\textbf{Data} &  \textbf{Oregon} & \textbf{Internet} & \textbf{Enron} & \textbf{Travian (Messages)} & \textbf{Travian (Trades)} & \textbf{hep-ph}\\
\midrule
\textbf{Min \# of nodes} & 10,670 & 2,948 & 101 & 1,373 & 964 &  12,711\\
\textbf{Max \# of nodes} & 11,174 & 6,477 & 137 & 2,100 & 1,336 & 34,401\\
\textbf{Min \# of edges} & 21,999 & 3,386& 1,432 &8,511 & 8,080 & 39,981\\
\textbf{Max \# of edges} & 23,409 & 13,233 & 5,015 &19,242 & 10,221 & 51,485\\
\textbf{\# of snapshots} & 9 & 733 & 12 &30 & 30 & 10\\
\bottomrule
\end{tabular}
\end{center}
\end{table*}

\textbf{AS-Oregon Graph ~\citep{leskovec2005graphs}}.  The dataset contains 9 graphs of Autonomous Systems (AS) peering information inferred from Oregon route-views between March 31, 2001 and May 26, 2001. These 9 graphs are different snapshots from the data with a minimum of 10,670 and maximum of 11,174 nodes. The number of edges ranges from 21,999 in the snapshot of April 07, 2001 to 23,409 in May 26, 2001. Figure~\ref{fig:changes2} shows the number of edges added and deleted, as well as the number of nodes involved in the changes for the AS-Oregon dataset.

\textbf{AS-Internet Routers Graph~\citep{leskovec2005graphs}}. Similar to AS-Oregon, this is a communication network of who-talks-to-whom from the Border Gateway Protocol logs of routers in the Internet. The dataset contains 733 daily snapshots for 785 days from November 8, 1997 to January 2, 2000. The number of nodes in the largest snapshot is 6,477 (with 13,233 edges).  Figure~\ref{fig:changes7} illustrates that the structure of the graph can change dramatically at each snapshot.

\textbf{Enron Email~\citep{sun2007graphscope}}. The Enron email network contains email message data from 150 users, mostly senior management of Enron Inc., from January 1999 to July 2002. Each email address is represented by an unique ID in the dataset, and each link corresponds to a message between the sender and the receiver. After a data refinement process, we simulate the network evolution via a series of 12 growing snapshots from January 2000 to December 2000. Enron network changes are shown in Figure~\ref{fig:changes3}.

\textbf{Travian}\footnote{ial.eecs.ucf.edu/travian.php}~\citep{hajibagheri2015conflict}
Travian is a popular browser-based real-time strategy game with more than 5 million users.  Players seek to improve their production capacity and construct military units in order to expand their territory through a combination of colonization and conquest.   Each game cycle lasts a fixed period (a few months) during which time the players vie to complete construction on one of the Wonders of the World. To do this, they form alliances of up to 60 members under a leader or a leadership team; in this article these alliances are used as the ground truth for evaluating the community detection procedure.  

Travian has an in-game messaging system (IGM) for player communication which was used to create our Messages network.  Each player can submit a request to trade a specific resource. If another player finds this request interesting, he/she can accept it and the trade will occur; this data was used to build the Trade network.   About 70\% of messages are exchanged between users in the same alliance (community) making it more predictive of community structure than the Trades network since only 30\% of edges in this network represent trades occurred between players within the same alliance. The structural changes in both Travian datasets are shown in Figures~\ref{fig:changes4} and~\ref{fig:changes5}.

\textbf{HEP-PH Citation Graph~\citep{leskovec2005graphs}}. The HEP–PH (high energy physics theory) citation graph from the e-print arXiv covers all the citations within a dataset of $n=29,555$ papers with $e= 352,807$ edges. If a paper $i$ cites paper $j$, the graph contains a directed edge from $i$ to $j$. If a paper cites, or is cited by, a paper outside the dataset, the graph does not contain any information about this. This data covers papers in the period from January 1993 to March 2002. There are ten snapshots where each snapshot includes data from twelve months except the last snapshot which only contains edges from the first three months of 2002 (Figure~\ref{fig:changes6}).
 
\subsection{Metrics}

We evaluated the performance of all methods using the following metrics.

\subsubsection{Normalized Mutual Information}
One way to measure the performance of a community detection algorithm is to determine how similar the partition delivered by the algorithm is to the desired partition, assuming ground truth information about the community membership exists. Out of several existing measures~\citep{fortunato2010community}, we selected the standard version of normalized mutual information (NMI)~\citep{danon2005comparing}, which is computed as follows:

\begin{eqnarray}\label{nmi1}
	\textbf{I}_{norm}\textbf{(X,Y)}=\frac{2I(X,Y)}{H(X)+H(Y)}, \nonumber \\
\end{eqnarray} 

\noindent
where $I(X,Y)$ is mutual information between two random variables $X$ and $Y$ (i.e. two community partitions)~\citep{mackay2003information}:

\begin{eqnarray}\label{nmi2}
\textbf{I(X,Y)}=\sum_x \sum_y P(x,y)\log\frac{P(x,y)}{P(x)P(y)}, \nonumber \\
\end{eqnarray} 
\noindent
Here $P(x)$ indicates the probability that $X=x$ and joint probability $P(x,y)$ equals to $P(X=x,Y=y)$. $H(X)$ and $H(Y)$ are the entropies of $X$ and $Y$, respectively.  NMI lies in the range [0,1], equaling one when two partitions $X$ and $Y$ are exactly identical and zero when they are totally independent.  Code to calculate NMI 
can be downloaded at: \url{https://sites.google.com/site/andrealancichinetti/software}.

\subsubsection{Modularity}
Modularity measures the difference between the number of intra-community edges for a given community partition vs.\ a random distribution of edges; it is the most popular qualitative measure in detecting communities in social networks.  However, it has been shown that modularity has drawbacks and becomes unreliable when networks are too sparse~\citep{fortunato2007resolution}.  It is also useful to examine the number of detected communities in conjunction with modularity to ensure that the algorithm is not being overly aggressive about combining small communities in order to maximize overall network modularity.

Standard modularity \textbf{Q} is usually defined as follows:

\begin{eqnarray}
\textbf{Q} = \frac{1}{2m} \sum_{ij}  [\textbf{A}_{ij} - \frac{d_id_j}{2m}]\delta_{c_i,c_j}
\end{eqnarray}
where \textbf{A}$_{ij}$ is an element of the adjacency matrix, $\delta_{ij}$ is the Kronecker delta symbol, and $c_i$ is the label of the community to which vertex $i$ is assigned. However, modularity can be slightly different for directed networks~\citep{leicht2008community} and can then be reformulated as:

\begin{eqnarray}
\textbf{Q} = \frac{1}{m} \sum_{ij}  [\textbf{A}_{ij} - \frac{d_i^{in} d_j^{out}}{m}]\delta_{c_i,c_j}
\end{eqnarray}
where \textbf{A}$_{ij}$ is defined in the conventional manner to be 1 if there is an edge from $i$ to $j$ and zero otherwise. Here, the probability of the existence of an edge from vertex $i$ to vertex $j$ has the probability $d_i^{in} d^{out}_j /m$, where $d_i^{in}$ and $d^{out}_j$ are the in- and out-degrees of the vertices respectively. 

\subsection{Evaluation}

In this article, we examine four research questions:
\begin{compactenum}
\item how does the gain function influence community detection performance?
\item does the initialization procedure affect the performance of dynamic community detection?
\item how does D-GT perform vs.\ the competitor methods?
\item in cases where the community membership of a small number of the agents is known, can it be used to improve D-GT's performance?
\end{compactenum}

\begin{table*}
\caption{Modularity evaluation metric on the six datasets; results are averaged over all snapshots.}
\label{tab:allmodularity}
\def\arraystretch{1.3}
\begin{center}
\begin{tabular}{lcccccc}
\toprule
\textbf{Algorithm/Dataset} &  \textbf{Oregon} & \textbf{Internet} & \textbf{Enron} & \textbf{Travian (Messages)} & \textbf{Travian (Trades)} & \textbf{hep-ph}\\
\midrule
\textbf{D-GT + Similarity} & 0.61$\pm$0.02 & \textit{0.59$\pm$0.02} & \textbf{0.50$\pm$0.02} & 0.45$\pm$0.03 & \textit{0.43$\pm$0.02} & \textbf{0.56$\pm$0.03}\\
\textbf{D-GTS + Similarity} & \textbf{0.63$\pm$0.02} & 0.57$\pm$0.04 & 0.48$\pm$0.02 & 0.44$\pm$0.02 & 0.42$\pm$0.01 & 0.55$\pm$0.02\\
\textbf{D-GTP + Similarity}& 0.59$\pm$0.02 & 0.58$\pm$0.05 & 0.47$\pm$0.03 & 0.44$\pm$0.03 & 0.41$\pm$0.02 & 0.51$\pm$0.04\\
\textbf{D-GT + Modularity} & 0.57$\pm$0.03 & 0.55$\pm$0.04 & 0.47$\pm$0.03 & 0.48$\pm$0.04 & 0.41$\pm$0.02 & 0.55$\pm$0.04\\
\textbf{D-GTS + Modularity} & \textbf{0.63$\pm$0.02} & 0.51$\pm$0.02 & 0.43$\pm$0.02 & \textit{0.49$\pm$0.03} & 0.40$\pm$0.03 & 0.54$\pm$0.02\\
\textbf{D-GTP + Modularity} & 0.59$\pm$0.03 & 0.48$\pm$0.03 & 0.39$\pm$0.02 & 0.47$\pm$0.03 & 0.42$\pm$0.02 & 0.52$\pm$0.04\\
\textbf{OSLOM} & 0.49$\pm$0.05 & 0.52$\pm$0.04 & 0.39$\pm$0.03 & 0.44$\pm$0.04 & 0.32$\pm$0.01 & 0.49$\pm$0.01\\
\textbf{LabelRankT} & 0.44$\pm$0.01 & 0.41$\pm$0.09 & 0.34$\pm$0.05 & 0.42$\pm$0.04 & 0.36$\pm$0.03 & 0.43$\pm$0.06\\
\textbf{iLCD} & 0.15$\pm$0.05 & 0.11$\pm$0.01 & 0.13$\pm$0.05 & 0.19$\pm$0.02 & 0.09$\pm$0.01 & 0.16$\pm$0.07\\
\textbf{InfoMap} & \textbf{0.63$\pm$0.04} & \textbf{0.61$\pm$0.01} & 0.49$\pm$0.08 & 0.51$\pm$0.02 & 0.43$\pm$0.04 & \textbf{0.56$\pm$0.04}\\
\textbf{Louvain} & 0.59$\pm$0.02 & 0.57$\pm$0.04 & 0.46$\pm$0.06 & \textbf{0.53$\pm$0.02} & \textbf{0.49$\pm$0.02} & 0.54$\pm$0.07\\
\bottomrule
\end{tabular}
\end{center}
\end{table*}


We analyze the performance of the D-GT variants vs. LabelRankT, iLCD, OSLOM, InfoMap, and Louvain on the two metrics, normalized mutual information (NMI) and modularity.  We hypothesize that D-GT with the modularity gain function will perform well on the modularity evaluation metric, since its stochastic search process essentially optimizes local modularity.  Also D-GT's initialization procedure will not necessarily help the modularity optimization process since it can be performed effectively without considering community evolution through time. This suggests that D-GTS is a good candidate for the modularity evaluation metric.  

However, in most real-world communities, prior community membership is a good predictor of future membership.  Thus we believe that D-GT is more effective at discovering the real community structure as measured by the normalized mutual information (NMI) evaluation metric.  Our previous work~\citep{alvari2011detecting} has shown that the neighborhood similarity function is a good gain function for recovering the true community structure so we hypothesize that D-GT (similarity) is the best choice for NMI.  

\begin{figure*}
 \centering
   {\includegraphics[width=1.1\columnwidth]{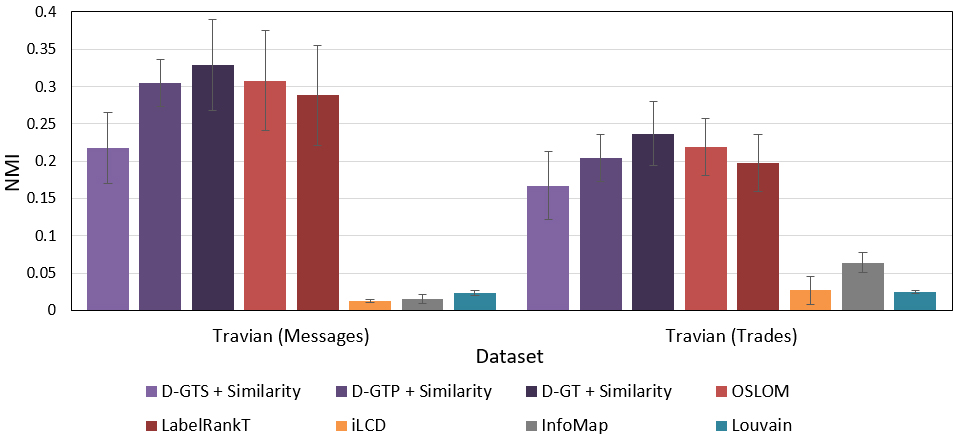}}
   \caption{Normalized mutual information (NMI) evaluation metric on the two Travian datasets with ground truth community membership information; results are averaged over all snapshots.   The variants of D-GT are colored in purple, LabelRankT (red), OSLOM (indian red), iLCD (yellow), InfoMap (gray) and Louvain (blue).}
  \label{fig:allsimilarity}
\end{figure*}

Figure~\ref{fig:allsimilarity} shows the average performance of all the D-GT variants with the similarity gain function vs.\ OSLOM, LabelRankT, iLCD, InfoMap and Louvain. Note that it is not possible to calculate the NMI performance on the AS-Internet, AS-Oregon, Enron, and hep-ph datasets since we don't have ground truth community structure; for the Travian datasets, we use the alliance membership to calculate the NMI.  D-GT (similarity) outperforms all other methods ($p<0.01$) on this metric.

\begin{figure*}
 \centering
   {\includegraphics[width=1.1\columnwidth]{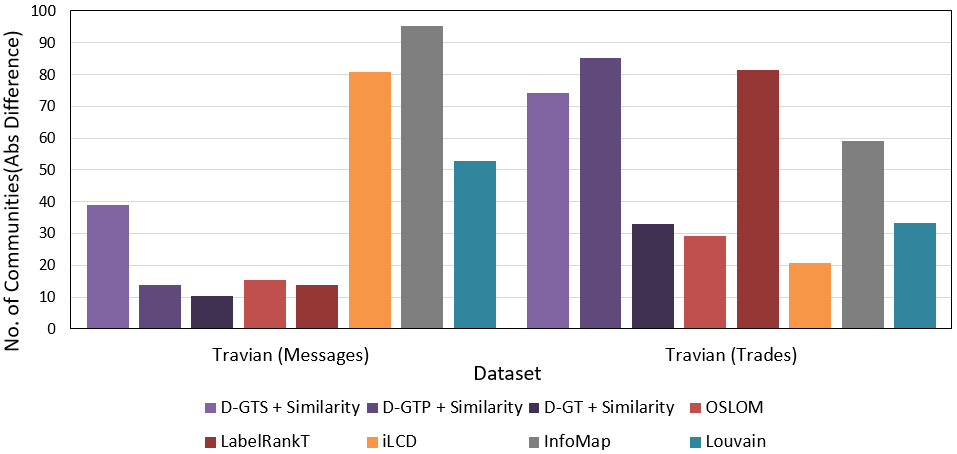}}
   \caption{Absolute difference between the predicted number of communities and the actual number for the two Travian datasets.  D-GT (with the similarity gain function) and OSLOM achieve the best performance overall at correctly predicting the number of alliances.}  (Since this serves a prediction error measurement, lower is better.)
  \label{fig:numcom}
\end{figure*}

In D-GT, each node's community membership vector is initialized as the superset of all communities that it has been a member of at any time step.   If most of the nodes have remained within the same communities, the correct structure is quickly discovered.  In cases where there are cyclic temporal patterns, there is clearly some value in considering earlier community assignments, 

As predicted it outperforms the more myopic D-GTS (that uses no prior information) and D-GTP (one previous snapshot); paired t-test comparisons on the two Travian datasets are significant at the $p<0.01$ level.  Figure~\ref{fig:numcom} shows that D-GT (similarity) is more successful than the other D-GT variants at correctly predicting the number of communities, as measured by summed absolute difference between predicted and actual community numbers (lower is better).   Note that it is possible to do acceptably well on the NMI metric while still incorrectly estimating the actual number of communities in the dataset.   OSLOM also performs well on both metrics (NMI and number of communities).

\begin{figure}
 \centering
 {\includegraphics[width=0.9\columnwidth]{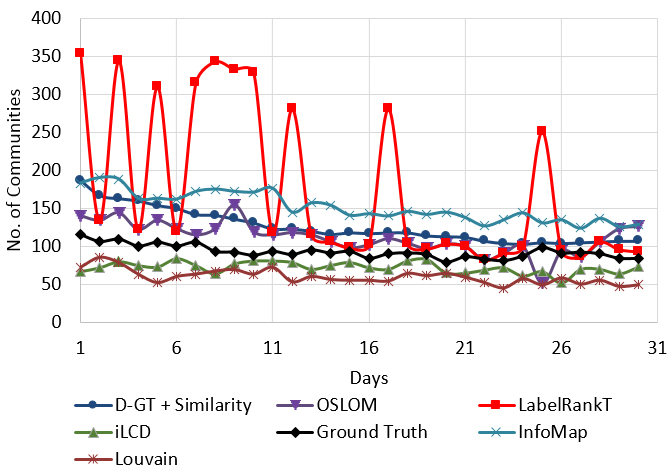}}
 \caption{Number of predicted communities vs. time for the Travian (Trades) dataset. LabelRankT's (red) predicted number of communities varies drastically between time steps, whereas all other algorithms make more consistent predictions. }
  \label{fig:numcomtimeseries}
\end{figure}

It is also useful to look at how the number of predicted communities varies between consecutive snapshots.  In most cases, the number of communities should remain relatively stable, since the structure of real-world communities rarely changes completely in short period of time.  This is definitely true in Travian, where the number of alliances changes relatively slowly. Figure~\ref{fig:numcomtimeseries} shows the number of predicted communities vs.\ time on the Travian (Trades) dataset; all of the methods make more consistent predictions over time than LabelRankT.  

Table~\ref{tab:allmodularity} shows the average performance of all the D-GT variants vs. OSLOM, LabelRankT, iLCD, InfoMap and Louvain at optimizing modularity.  Bold font shows the absolute best performing algorithm, with italics marking the best performing D-GT variant.  All D-GT variants are competitive at optimizing modularity but none are exceptional.  They outperform the other dynamic algorithms (iLCD, LabelRankT) but rarely 
the static community detection algorithms (Louvain and InfoMap).  This is unsurprising since the modularity metric does not intrinsically reward preserving continuity between snapshots.  D-GT (similarity) continues to perform well, as does D-GTS (modularity).

\begin{figure}
 \centering
 {\includegraphics[width=0.9\columnwidth]{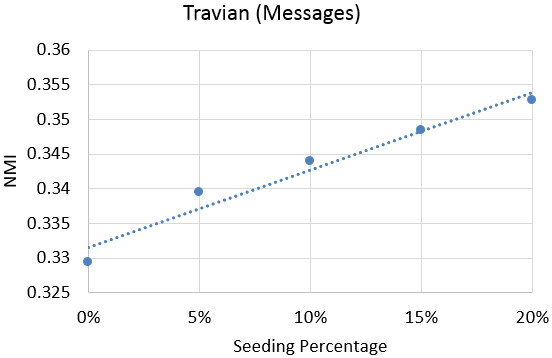}}
 \caption{D-GTG NMI vs.\ seed group size on Travian (Messages)}
  \label{fig:dgtgmessages}
\end{figure}

\begin{figure}
 \centering
 {\includegraphics[width=0.9\columnwidth]{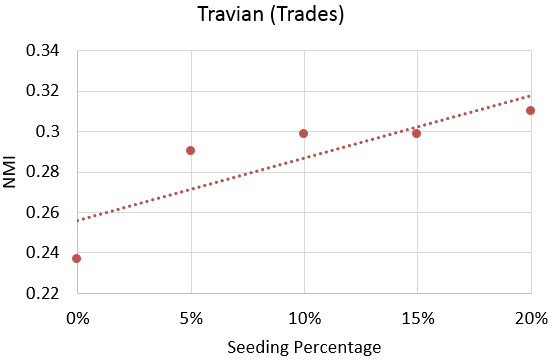}}
 \caption{D-GTG NMI vs.\ seed group size on Travian (Trades)}
  \label{fig:dgtgtrades}
\end{figure}

In some scenarios, it is plausible that the community membership of a small number of agents is known in advance, and the community detection procedure should leverage this information.  For instance, MMOG game alliances often have a leadership council that is publicly known.  To handle this problem, we developed a variant (D-GTG: D-GT with passing Ground Truth).  Fig.~\ref{fig:dgtgmessages} shows the performance improvements from increasing the size of the seed groups from 0--20\% of the total number of agents for the Travian (Messages) dataset, and Fig.~\ref{fig:dgtgtrades} shows the performance increase for Travian (Trades).  Note that extracting community membership information from the network structure of Travian (Trades) is a difficult problem because only 30\% of the edges in Travian (Trades) occur between players within the same alliance (community).  Also the dataset has a high number of isolated nodes; about 50\% of the nodes do not belong to any alliance.

\begin{table}[t]
	\caption{Running time (sec) of all algorithms on the AS-Internet, AS-Oregon, and hep-ph datasets}
	\centering
	\label{tb:runtime}
	\begin{tabular}{|l|c|c|c|}
		\hline
		\textbf{Algorithm} & \textbf{Internet} & \textbf{Oregon} & \textbf{hep-ph}\\\hline
		\textbf{D-GT} & 610 & 45 & 1,140\\
		\textbf{LabelRankT} & 840 & 90 & 1,250\\
		\textbf{iLCD} & 750 & 80 & 1,000\\
		\textbf{OSLOM} & 590 & \textbf{35} & 900\\
        \textbf{InfoMap} & \textbf{240} & 42 & \textbf{620}\\
        \textbf{Louvain} & 380 & 60 & 730\\
        \hline
	\end{tabular}
\end{table}

Table~\ref{tb:runtime} shows the running time of all algorithms on our largest datasets, the AS-Internet dataset (with the highest number of snapshots) and hep-ph (with the highest number of nodes). Overall, the running times provided by all algorithms were quite reasonable, as there are 733 snapshots in the Internet dataset and over 10,000 and 30,000 nodes in Oregon and hep-ph respectively. InfoMap is the fastest algorithm on two datasets (Internet and hep-ph) and performs almost as well as OSLOM on third one (Oregon).  All of the algorithms are sufficiently fast to run on large datasets.
\section{Conclusion}
\label{sec:conclusion}

This article analyzes the performance of our game theoretic community detection algorithm, D-GT, on dynamic social networks.  These social networks are very common in massively multiplayer online games, such as Travian, where players self-organize into rapidly changing guilds and alliances.  We show that D-GT's initialization procedure in combination with the similarity gain function is very effective at recovering the true community structure of the network, both in terms of predicting the number of communities and the players' community membership vectors. It outperforms other dynamic community detection methods including LabelRankT, iLCD, and OSLOM. In cases where the communities of a small number of players (e.g. the guild leadership) is known D-GT can leverage the information to improve the NMI performance.  When simply optimizing modularity, considering earlier community membership is less important and static community detection algorithms (InfoMap and Louvain) perform well at this task, but all variants of D-GT offer competitive performance.

One nice aspect of D-GT is that it is an easily extensible framework.  In future work we plan to experiment with other utility functions, loss functions, and update rules; for instance, Q-learning could be used as the update rule for the agents instead of the community membership game.  We also believe that it is feasible to reduce the runtime of D-GT by creating approximate versions of the gain function that can be calculated based on local edge updates.

\section*{Acknowledgments}
We would like to thank Drs.\ Nitin Agarwal and Rolf T.\ Wigand at University of Arkansas for providing the Travian dataset.

\end{document}